\documentstyle[aps,preprint,amssymb]{revtex}
\begin{document}
\begin{titlepage}
\begin{center}
\makebox[\textwidth][r]{SNUTP 96-9}
\makebox[\textwidth][r]{UFIFT-HEP-96-3} 
\vskip 0.35in
{{\Large \bf 
RG analysis of magnetic catalysis in Dynamical symmetry breaking
}}
\end{center}
\begin{center}
\par \vskip .1in 
\noindent Deog Ki Hong\footnote{\tt dkhong@@hyowon.cc.pusan.ac.kr}$^{,a,b}$, 
Youngman Kim \footnote{\tt han@@hepth.hanyang.ac.kr}$^{,c}$
and Sang-Jin Sin\footnote{\tt sjs@@hepth.hanyang.ac.kr}$^{,c}$
\end{center}
\begin{center}
{\it $^a$Department of Physics, Pusan National 
University\footnote{\it permant address}\\  
Pusan 609-735, Korea\\
$^b$Institute of Fundamental Theory\\ 
Department of Physics,  University of Florida\\ 
Gainesville, FL32611, U.S.A.\\
$^c$Department of Physics, Hanyang University\\ Seoul, Korea\\ 
} 
\par \vskip .1in 
\noindent
\end{center}
\begin{abstract}
We perform the renormalization group analysis on the dynamical symmetry 
breaking under strong external magnetic field, studied recently 
by Gusynin, Miransky and Shovkovy. We find that any attractive 
four-Fermi interaction becomes strong in the low energy, thus 
leading to dynamical symmetry breaking. When the four-Fermi interaction 
is absent, the $\beta$-function for the electromagnetic coupling 
vanishes in the leading order in $1/N$. By solving 
the Schwinger-Dyson equation for the fermion propagator, 
we show that in $1/N$ expansion, 
for any electromagnetic coupling, dynamical symmetry breaking occurs 
due to the presence of Landau energy gap by the external magnetic field. 
\vskip 0.2in
\noindent
PACS numbers: 11.30.Rd, 11.10.Gh, 12.20.Ds, 11.15.Pg
\end{abstract}
\end{titlepage}
Recently dynamical symmetry breaking in the presence of the external 
magnetic field has attracted much interest.
Gusynin, Miransky and Shovkovy \cite{gusynin} have shown 
that constant magnetic field
acts as a strong catalyst of dymamical symmetry breaking, 
which leads to fermion mass
generation in 2+1 as well as in 3+1 dimension. 
They also have shown that similar
phenomena happen in 3+1 QED without extra four fermion coupling \cite{gusynin}.
In a recent article \cite{dunne} Dunne and Hall discussed the problem in the
context of a non-uniform magnetic field in 2+1 dimesnsion.

In this paper we perform
the renormalization group analysis and show that the 4-fermi interaction
becomes strong in the infrared and leads to instability that causes
dynamical symmetry breaking.
For QED under strong external magnetic fields, 
we do not have a RG interpretation for dynamical mass generation, 
because the electromagnetic coupling is not running.   
Thus, one has to rely on  
nonperturbative methods to see dynamical mass generation. 
We solve the 
Schwinger-Dyson equations for the fermion propagator in the $1/N$ expansion 
and find nontrivial and energetically preferable solutions. By 
comparing the solutions with operator product expansion for the 
fermion two-point function, we argue that 
for any small electromagnetic coupling fermions get 
dynamical mass. 
This result is consistent with the analysis 
of Bethe-Salpeter equation for fermion and anti-fermion bound state 
done by Gusynin {\it et al.} \cite{gusynin}.
  
We first consider a free fermion under constant magnetic field in 
$z$ direction, ${\bf B}=B\hat z$.
The Lagrangian is given by
\begin{eqnarray}
{\cal L}_0=\bar\Psi i\not\!\! D \Psi
\end{eqnarray}
where $\not\!\!\! D=\gamma^{\mu}(\partial_{\mu}-ie A^{\rm ext}_{\mu})$
and $\Psi$ is a massless fermion.
We choose the symmetric gauge ${\bf A}=(-{B\over2}y,{B\over2}x,0)$
and solve the eigenvalue equation, 
\begin{eqnarray}
[ \vec{\alpha}\cdot({\bf p}- e {\bf A}) ]\Psi=E\Psi
\end{eqnarray}
to get the spectrum and the basis of this system.
The eigenvalues are indexed by collective index $A=(\alpha,\beta,n,k_z)$
and given by 
\begin{eqnarray}
E_A=\alpha\sqrt{k_z^2+ 2\left|eB\right|n}
\end{eqnarray}
where $\alpha=\pm$ denotes the sign of the energy, $\beta=\pm{1\over2}$
is the spin component along the magnetic field, and the quantum number $n$ 
is given by 
\begin{eqnarray}
2n=2n_r+1+|m_L|-{\rm sign}(eB) (m_L+2\beta). 
\end{eqnarray}
$n$ is a nonnegative integer that labels the Landau level.
Here $n_r$ is the number of nodes of radial eigenfuction, $m_L$ is the angular
momentum of the eigen function.
The eigen function is 
\begin{eqnarray} 
U_A=N_A e^{ik_z z}e^{im_L\phi}r^{|m_L|}L_{n_r}^{|m_L|}
\left({1\over2}|eB|r^2\right) u_{\alpha,\beta}, 
\end{eqnarray}
where $N_A $ is the nomalization, $L_n^r(x) $ is the associated Laguerre 
polynomial, and $u_{\alpha,\beta}=\chi_\alpha\otimes\eta_\beta$ 
is the eigenvector of $\sigma_3\otimes\sigma_3$ where two $\sigma_3$
 correspond to the energy and the spin. 
It is a spinor in the Lorentz frame in 
which the magnetic field is specified. This basis forms an orthonormal system.
The eigenfunction expansion
\begin{eqnarray} 
\Psi(x,t)=\int\!\!\sum_A \psi_A(t)U_A(x),
\end{eqnarray}
yields  
\begin{eqnarray} 
S_0= \int\!\!\sum_{A,t}\psi_A(t)^\dagger (i\partial_t- E_A)\psi_A(t)
\end{eqnarray}
Finally we take the Fourier transform in $t$ to get 
\begin{eqnarray} 
S_0=\int d^4x  \bar\Psi i\not\!\! D \Psi= \int\!\!\sum_{A,\omega}\psi_A^\dagger (\omega- E_A)
\psi_A(\omega)
\end{eqnarray}
where 
\begin{eqnarray}
\int\!\!\sum_{A,\omega}
=\sum_{\alpha,\beta,n_r,m_L}\int{dk_z\over2\pi }{d\omega\over2\pi}
\end{eqnarray}
We now determine the scaling dimesions of various modes.
Notice that $n=0$, the lowest mode, has a different scaling 
property  from the rest of the modes.
Under the scaling, 
\begin{eqnarray}
k_z\to s k_z, \omega\to s\omega,\;\; {\rm with}\;\; s<1 
\end{eqnarray}
we require that the kinetic term be invariant. To meet this, 
$\psi_0(k_z,\omega)$ must have scaling dimension $-3/2$, 
while $\psi_{n>0}(k_z,\omega)$ must have $-1$, since the Landau gap 
should not scale.
Namely, 
\begin{eqnarray}
\psi_0(k_z,\omega)&\to s^{-3/2}\psi_0(\omega,k_z)\\ 
\psi_n(k_z,\omega)&\to s^{-1}\psi_n(\omega,k_z) \quad{\rm for}\;\;\; n>0. 
\end{eqnarray} 
We see that the $n=0$ mode is more relevant than the other modes.
This is a manifestation of the dominance of the lowest Landau level 
(LLL) or 
decoupling of the fermions at higher Landau levels.
We now determine the scaling dimension of the interaction.
Again, we split the interaction term $S_{int}$ into two
parts; one contains only zero modes, 
the other contains at least one $n\ne0$
mode; $S_{int}=S_{int}^0+S'_{int}$, where 
\begin{eqnarray}
S_{int}^0=\int\prod_{i=1}^4dk_i d\omega_i \delta(\sum_i k_i)
\delta(\sum_i\omega_i)
\psi^\dagger_{0,A}\psi_{0,B}C_{ABCD}\psi^\dagger_{0,C}\psi_{0,D} 
\end{eqnarray}
with 
\begin{eqnarray}
C_{ABCD}=\int dxU^{\dagger}_A(x)U_B(x)U^{\dagger}_C(x)U_D(x)
\end{eqnarray}
where each of $A,B,C,D$ contains the LLL index $n=0$.
The dimensionality of $S_{int}^0$ is $4(2+(-3/2)\times 2\times 2)-1-1=0$.
Since the zeroth mode is the most relevant term, all the rest terms have  
positive scaling dimension therefore they are irrelevant interactions.
Therefore the transverse directions are completely decoupled from the low 
energy dynamics. The interactions are at most marginal at tree level.
This is the proof of the dimensional reduction which was also noticed 
in \cite{gusynin} from the structure of the Schwinger propagator.
So to settle the issue whether there is any relevant interaction that leads
phase transition, we have to go to the one loop $\beta$ function. 
Unless there is a
special reason, the one loop $\beta$ function does not vanish in general.  
The crucial factor then is its sign. 

Consider NJL model in $(3+1)$D with external electromagnetic field,
$A^{\rm ext}_{\mu}(x)$, described by Lagrangian density,
\begin{eqnarray}
{\cal L}=\bar\Psi i\not\!\! D \Psi+{G\over2}
        \left[ \left(\bar\Psi \Psi \right)^2+
        \left(\bar\Psi i \gamma^5 \Psi\right)^2\right]
\end{eqnarray}
where $\not\!\! D=\gamma^{\mu}(\partial_{\mu}-ie A^{\rm ext}_{\mu})$
and $\Psi$ is a massless
$N$-column 4-component spinor.
Let $\Lambda$ be 
some scale of the system under study. 
We choose it to be less than the Landau level gap 
$\sqrt{\left|eB\right|}$. 
we decompose the field $\Psi$ into slow and fast components
\begin{eqnarray}
\Psi=\Psi_s+\Psi_f,  
\end{eqnarray}
where 
\begin{eqnarray}
\Psi_s=\sum_{|k_z|<s\Lambda}e^{ik_z z}\Psi(x,y,k_z),\quad\quad
\Psi_f=\sum_{s\Lambda<|k_z|<\Lambda}e^{ik_zz}\Psi(x,y,k_z).
\end{eqnarray}
Since the transverse momentum 
should not scale 
we do not split it into slow and fast. 
We integrate the fast modes out and see how much change is  
induced in the action of slow part.
For example out of 
$ (G/2)^2[({\bar\Psi_s}+{\bar\Psi}_f)(\Psi_s+\Psi_f)]^4$ 
we are looking for the coefficient of $({\bar\Psi_s}\Psi_s)^2$.
We only get $2!\times2!$ times the fish diagram. Since the calculation is
straightforward we only mention the structure of the propagator.

As derived by Schwinger \cite{schwinger}, 
the fermion propagator in a constant external magnetic field is 
\begin{eqnarray}
S(x,y)={\tilde S}(x-y) \exp\left[{ie\over 2}(x-y)^{\mu} A^{\rm ext}_{\mu}
(x+y )\right]
\end{eqnarray} 
We take the magnetic field in $z$ direction. 
The Fourier transform of $\tilde S$ is given as 
\begin{eqnarray}
\tilde S(k)=i e^{-\vec k_{\perp}^2/\left|eB\right|} 
\sum_{n=0}^{\infty}(-1)^n{D_n(eB,k)\over 
k_0^2-k_3^2-2\left|eB\right|n}
\end{eqnarray}
with 
\begin{eqnarray}
D_n(eB,k)&=(k^0\gamma^0-k^3\gamma^3)
\Big[ \left( 1-i\gamma^1\gamma^2 {\rm sign }(eB)\right)
L_n\left({2k_{\perp}^2\over \left|eB\right|}\right)\\
-\Big( 1+i&\gamma^1\gamma^2 {\rm sign }(eB)\Big)
L_{n-1}\left({2k_{\perp}^2\over \left|eB\right|}\right) \Big]
+4\vec k_{\perp}\cdot\vec\gamma_{\perp} 
L_{n-1}^1\left({2k_{\perp}^2\over \left|eB\right|}\right),
\end{eqnarray}
where $L^{\alpha}_n$ are the generalized Laguerre polynomials
and $L_n\equiv L_n^0$, ${L^{\alpha}}_{-1}=0$. $k_{\perp}$ is momentum 
perpendicular to the external magnetic field. 
As we have shown above, only LLL fluctuation contributes to the low energy
dynamics so we can use the propagator that is projected 
to LLL instead of the Schwinger's full propagator.
\begin{eqnarray}
\tilde S(k)=i e^{-\vec k_{\perp}^2/\left|eB\right|}
{\not \! k_{\shortparallel}\over k_{\shortparallel}^2}P_-
+O \left({k_{\shortparallel}^2\over \left|eB\right|} \right),
\end{eqnarray}
where $k_{\shortparallel}=k-k_{\perp}$ and 
$P_-=1-i\gamma^1\gamma^2{\rm sign}(eB)$ is the projection
operator to project out fermions of spin parallel to the external magnetic
field if $e>0$ or antiparallel if $e<0$.
The one loop correction to the four-Fermi interaction is 
\begin{eqnarray}
i\delta G&= ( {G\over 2})^2N\int_{s\Lambda<|k|<\Lambda} 
             {\rm Tr}\left[ \tilde S(k) \tilde S(k)
             +\tilde S(k) i\gamma^5\tilde S(k)i\gamma^5\right] \\
           &= 2G^2N\int_{s\Lambda<|k_{\shortparallel}|<\Lambda} 
                e^{-k_{\perp}^2/\left|eB\right|-
        k_{\perp}^2/\left|eB\right|} {{\rm Tr}\left[\not\!k_{\shortparallel}
        \not\!k_{\shortparallel}\right]\over 
        k_{\shortparallel}^2  k_{\shortparallel}^2}P_-^2\\ 
&= {iN\over 4\pi^2 } \left|eB\right| G^2 \ln{1\over s} 
\end{eqnarray}
If we set $g=|eB|G$ with dimensionless coupling $g$, then
$\delta g=|eB|\delta G $. 
Hence the above equation can be written as 
\begin{eqnarray} 
\beta(g)=s{\partial g\over \partial s}=-{N\over 4\pi^2}g^2
\label{rge}
\end{eqnarray}

The dynamical mass generated in the infrared can be determined 
by integrating the renormalization group equation Eq. (\ref{rge}). We find  
\begin{eqnarray}
m_{\rm dyn}\sim \sqrt{\left|eB\right|}e^{-{4\pi^2\over N g 
\left( \left|eB\right|\right)}}
\label{dynamass}
\end{eqnarray}
The magnitude of dynamical mass is determined by the strength of the 
four-Fermi coupling at scale $\sqrt{\left|eB\right|}$.
This result agrees with the vacuum energy calculation
of Gusynin {\it et al.} 

Now, let us consider the case when four-Fermi interaction is absent, 
namely, pure QED, in 
which case the symmetry of the system is enlarged;
\begin{eqnarray}
SU(N)\longrightarrow SU(N)_L\times SU(N)_R
\end{eqnarray}
In general, the electromagnetic coupling is weak in the  
IR region. Therefore nonperturbative effects like dynamical symmetry 
breaking do not occur unless QED is in the strong phase \cite{strongqed}.  
But, under a strong external magnetic field, since the excitations of fermions 
occur only in the magnetic field direction, fermions may interact strongly. 
To see this one has to 
rely on another nonperturbative method. 
Here we use $1/N$ expansion, keeping $e^2N=\alpha$ finite when 
$N\to \infty$. 
Then, for the photon propagator, we have to sum all bubble diagrams.  
The one loop vacuum polarization at low energy  is  
\begin{eqnarray}
\Pi^{\alpha\beta}(p)&=-\left({-ie}\right)^2\cdot  
N\int_k {\rm Tr} 
\left[\tilde S(k)\gamma^{\alpha}\tilde S(p+k)\gamma^{\beta} \right]\\
&=+\alpha\int_k e^{-{\vec k_{\perp}^2 
\over \left|eB\right|}-{(\vec k_{\perp}+\vec p_{\perp})^2
\over \left|eB\right|}}
{\rm Tr} \left[
{\not\! k_{\shortparallel}\over k_{\shortparallel}^2}
P_-
\gamma^{\alpha} {\not \! k_{\shortparallel}+\not\! p_{\shortparallel}
\over (k_{\shortparallel}+p_{\shortparallel}) ^2}
P_-\gamma^{\beta}\right]\\
&={i\alpha\over 2\pi^2} \left|eB\right|
e^{-p_{\perp}^2/2\left|eB\right|}
\left( g^{\alpha\beta}_{\shortparallel}-
p_{\shortparallel}^{\alpha}p_{\shortparallel}^{\beta}
/p_{\shortparallel}^2\right),
\end{eqnarray}
where $g_{\shortparallel}^{\alpha\beta}=g^{\alpha\beta}$ for 
$\alpha,\beta=0,3$ otherwise zero.  
The photon propagator in $1/N$ expansion is then in Landau gauge  
\begin{eqnarray}
D^{\alpha\beta}(p)=-i{g^{\alpha\beta}-p^{\alpha}p^{\beta}/p^2\over p^2}
-{\Pi^{\alpha\beta}(p)\over p^2\left(p^2+i\Pi(p)\right)},
\end{eqnarray}
where 
\begin{eqnarray}
\Pi(p)={i\alpha\over 2\pi^2}\left|eB\right| e^{-p_{\perp}^2/2\left|eB\right|}.
\end{eqnarray}
We see that the singularity of the photon propagator in the 
parallel direction is 
shifted from $p=0$ to nonzero $p$. Photon develops mass along 
the parallel direction 
$m^2= {\alpha\over 2\pi^2} \left|eB\right|$, 
while it remains massless along the perpendicular direction. 
The Coulomb potential between two static fermions is  
\begin{eqnarray}
V(r)=\int_{-\infty}^{\infty}
dx_0\int {{d^4p}\over (2\pi)^4}iD^{00}(p)e^{ip\cdot x}
\cong {e^{-mr}\over 4\pi r}.
\end{eqnarray}
Since the vacuum polarization tensor $\Pi^{\alpha\beta}(p)$ is finite, 
the electromagnetic coupling is not renormalized at the leading 
order in $1/N$ expansion. 

The simplest order parameter for symmetry breaking is 
the condensate of fermion bilinear;
\begin{eqnarray}
\left<\Psi\bar\Psi\right>&=\lim_{y\rightarrow x}\left|x-y\right|^{\gamma_m}
\left<\Psi(x)\bar\Psi(y)\right>\\
&=\left<\psi_0\bar\psi_0\right>,
\end{eqnarray}
where $\gamma_m$ is the anomalous mass dimension of $\Psi\bar\Psi(x)$ and 
in the second line we used the fact that fermions at higher Landau levels 
decouple.
One can extract the value for the order parameter from the operator 
product expansion of the fermion propagator \cite{georgi}. 
Solving the Schwinger-Dyson equations, we determine the fermion propagator. 
For energy less than the Landau gap, $\sqrt{\left|eB\right|}$, 
the full fermion propagator will be of following form;
\begin{eqnarray}
\tilde S(k)=ie^{-k_{\perp}^2/\left|eB\right|}
{1\over Z(k_{\shortparallel})\not\!k_{\shortparallel}-
\Sigma(k_{\shortparallel})}P_-.
\end{eqnarray}
Note that the fermion self energy $Z$ and $\Sigma$ are independent of 
$k_{\perp}$, since the on-shell condition has to be 
\begin{eqnarray}
k_{\shortparallel}^2+m_{\rm phy}^2=0,
\end{eqnarray} 
where $m_{\rm phy}$ is the physical mass.
Schwinger-Dyson equation for the fermion propagator  
in the leading order in $1/N$ is 
\begin{eqnarray}
\left[\left(Z(p_{\shortparallel})-1\right)\not\!p_{\shortparallel}
-\Sigma(p_{\shortparallel})\right] P_-
={\alpha\over N}\int_k e^{-k_{\perp}^2/\left|eB\right|}
\gamma^{\mu}{D_{\mu\nu}(p_{\shortparallel}-k)\over 
Z(k_{\shortparallel})\not\!k_{\shortparallel}
-\Sigma(k_{\shortparallel})} P_-\gamma^{\nu}.
\end{eqnarray}
At leading order, we can take $Z(p_{\shortparallel})=1$. 
Then the (chirally non-invariant) fermion self energy is, 
after Wick rotation 
and taking trace over Dirac matrices, 
\begin{eqnarray}
\Sigma(p_{\shortparallel})={\alpha\over N}\int{d^4k\over (2\pi)^4}
e^{-k_{\perp}^2/\left|eB\right|}
{\Sigma(k_{\shortparallel})\over k_{\shortparallel}^2+
\Sigma(k_{\shortparallel})^2}{1\over (p_{\shortparallel}-
k_{\shortparallel})^2+k_{\perp}^2}.
\label{sd}     
\end{eqnarray}
After integrating over $k_{\perp}$ and isolating the logarithmic 
divergence at $(p_{\shortparallel}- k_{\shortparallel})^2=0$,  
we expand the r.h.s. of Eq. (\ref{sd}) 
in powers of $(p_{\shortparallel}- k_{\shortparallel})^2/\left|eB\right|$ 
and get 
\begin{eqnarray}
\Sigma(p_{\shortparallel})={\alpha\over N}\int_{k_{\shortparallel}}
{\Sigma(k_{\shortparallel})\over k_{\shortparallel}^2+
\Sigma(k_{\shortparallel})^2}
\left[ -{1\over4\pi}\ln{(p_{\shortparallel}- k_{\shortparallel})^2\over
\left|eB\right|}-0.063+\cdots \right],
\label{sd1}
\end{eqnarray} 
where the ellipsis denotes terms of order of 
$ (p_{\shortparallel}- k_{\shortparallel})^2/
\left|eB\right|$ or higher which are negligible. 
(From now on, we let $p=\left|p_{\shortparallel}\right|$ and 
$k=\left|k_{\shortparallel}\right|$.) 
After differentiating Eq. (\ref{sd1}) with respect to $p$,   
we integrate over the direction of $k_{\shortparallel}$ to get 
\begin{eqnarray}
{\partial\over \partial p}\Sigma(p)=-{\alpha\over 2\pi^2 N}
\int_0^pdk {k\over p} {\Sigma(k)\over k^2+\Sigma(k)^2}.
\label{sd2}
\end{eqnarray}
Multiplying by $p$ and differentiating once again, we obtain 
\begin{eqnarray}
p \Sigma^{\prime\prime}+\Sigma^{\prime}+r{p\Sigma\over p^2+\Sigma^2}=0,
\label{diff}
\end{eqnarray}
where  the prime denotes differentiation with respect to $p$ and 
we have defined 
\begin{eqnarray}
r={\alpha\over 2\pi^2 N}.
\end{eqnarray}
An infrared boundary condition is given by Eq. (\ref{sd2});
\begin{eqnarray}
\lim_{p\rightarrow0}p\Sigma^{\prime}(p)=0.
\label{irbc}
\end{eqnarray}
One can easily find the soulutions to the second-order, nonlinear 
differential equation (\ref{diff}) in two asymptotic regions. 
For $p\ll\Sigma(p)$, 
\begin{eqnarray}
\Sigma(p)\simeq m_C+\lambda \ln\left({p\over \mu}\right).
\label{irsol}
\end{eqnarray}
But, because of the IR boundary condition (\ref{irbc}), 
$\lambda=0$ and $\Sigma(p)\simeq m_C$ for $p\ll \Sigma(p)$. For 
$p\gg\Sigma(p)$,
\begin{eqnarray}
\Sigma(p)\simeq m\left({\mu\over p}\right)^{\epsilon}
+\kappa \left({\mu\over p}\right)^{-\epsilon},
\label{uvsol}
\end{eqnarray}
where $\epsilon=i\sqrt{r}$. The parameter $\mu$, with mass dimension 
corresponds to the renormalization point. 
By using same analysis done by Cohen and Georgi \cite{georgi}, 
one can show that there are solutions which connect two asymptotic 
solutions. 

We let 
\begin{eqnarray}
X(t)\equiv \Sigma(p)/p, \quad t\equiv \ln(p/\mu).
\end{eqnarray}
Then Eq. (\ref{diff}) becomes
\begin{eqnarray}
\ddot{X}+2\dot{X}=-{d\over dX}V(X),
\end{eqnarray}
where 
\begin{eqnarray}
V(X)={1\over 2}\left[ X^2+r\ln(1+X^2)\right].
\end{eqnarray}
This is the equation for a particle of unit mass moving in a potential 
$V$ under a friction proportional to veleocity. 
The IR boundary condition implies that as $t\to -\infty$ the particle is 
sitting at 
\begin{eqnarray}
X(t)\to {m_C\over \mu}e^{-t}.
\end{eqnarray} 
Since, for large $X$, $V\simeq {1\over2}X^2$, the particle moves 
toward $X=0$, critically damped by the friction. But, near $X=0$, the 
particle is underdamped. Therefore the particle sitting at $X(-\infty)$ 
will eventually get to $X=0$ in a finite time and then oscillate 
around $X=0$, corresponding to the asymptotic solutions 
(\ref{uvsol}). 
We see therefore there must exist solutions which interpolate 
the IR and the UV solutions. 
Futhermore, since the the vacuum energy at the leading order  
in $1/N$ is  
\begin{eqnarray}
V(\Sigma)={\left|eB\right|\over 4\pi}\int{d^2k_{\shortparallel}\over (2\pi)^2}
\left[ -2\ln(k_{\shortparallel}^2+\Sigma^2)
+2{\Sigma^2\over k_{\shortparallel}^2+\Sigma^2}\right],
\label{vac} 
\end{eqnarray}
such solutions have lower vacuum energy than the trivial 
solution, $\Sigma=0$. 

The operator product expansion for the fermions in the lowest Landau 
level has the following form:
\begin{eqnarray}
\lim_{-p^2\rightarrow\infty}\left<\psi_0\bar\psi_0(p)\right>&=
{1\over \not\!p}+{\Sigma(p)\over p^2}+\cdots\\
&={1\over \not\!p}+{m\over p^2}\left({\mu\over p}\right)^{\epsilon}
+{\left<\psi_0\bar\psi_0\right>\over p^2}
\left({\mu\over p}\right)^{-\epsilon}+\cdots, 
\end{eqnarray}
where $p^2=w^2-k_z^2$. 
(Note that two operator $m$ and $\psi_0\bar\psi_0(x)$ have same 
mass dimension but opposite anomalous mass dimension.) 
But, since the nontrivial solutions to the gap equation 
(\ref{diff}) have oscillatory behavior in deep UV region 
(due to imaginary anomalous mass dimension),  
one cannot tell from the operator product expansion of 
the fermion propagator which is the renormalized mass and which is 
the fermion condensate. But, as a signal for the dynamical symmetry breaking,  
we see the renormalized mass and the fermion condensate 
coalesce for any value of electromagnetic coupling $\alpha$. 
Therefore, as a general rule, we have dynamical symmetry breaking, 
$\left<\Psi\bar\Psi\right>\ne0$ in QED in a strong magnetic field.

In conclusion, we analyzed dynamical symmetry breaking under  
external magnetic field in the Wilsonian renormalization picture. 
We found that any attractive four-Fermi interaction is enhanced in 
the infrared, leading to dynamical mass generation for fermions. 
When the four-Fermi interaction is absent, the $\beta$-function 
for the electromagnetic coupling is zero in the leading order in $1/N$. 
By solving the Schwinger-Dyson equation, we found that any small 
electromagnetic coupling leads to dynamical symmetry breaking, just 
because at low energy ($E<\sqrt{\left|eB\right|}$) fermions at high Landau 
levels decouple and fermions at lowest Landau level are (1+1)-dimensional 
and photons are propagating in (3+1)-dimensions.  
It will be of great interest to extend this result to the non-homogeneous 
magnetic field context. As a prelimanary step to this direction,
one can study the small deviation from the constant magnetic field.
In this case one can easily show that this deviation contains at least one power of momentun and it is irrelevant
interaction term, therefore does not chage the ground state qualitatively.
This background deviation, of course, is different from the 
quantum fluctuation of the photon field whose effect is shown to be relevant 
in this paper indirectly.
\vskip .1in
\noindent
{\bf Acknowledgments}
\vskip .1in
We thank P. Ramond and H.K. Lee for discussions.  
D.K.H. is grateful to the members of the IFT of 
university of Florida for their hospitality during his stay at IFT.  
This work was supported in part by the KOSEF through SRC program of SNU-CTP 
and also by Basic Science Research Program, Ministry of Education, 
1995 (BSRI-95-2413 for D.K.H, and BSRI-95-2441 for S.J.S). 
The work of S.J.S. was also supported in part  
by the KOSEF (95-1400-04-01-3) and by Hanyang University.

\pagebreak




\begin{thebibliography}{99}

\bibitem{gusynin}V.P. Gusynin, V.A. Miransky, and I.A. Shovkovy, 
Phys. Rev. Lett. {\bf 73} (1994) 3499; Phys. Rev. {\bf D 52} 
(1995) 4718; hep-ph/9509320. 

\bibitem{dunne} G. Dunne and T. Hall, preprint 
"Inhomogenious condensates in planar QED"; hep-th/9511192

\bibitem{schwinger}J. Schwinger, Phys. Rev. {\bf 82} (1951) 664; 
J.H. Lowenstein and J.A. Swieca, Ann. Phys. (N.Y.) {\bf 68} 
(1971) 172.

\bibitem{strongqed}P.I. Fomin, V.P. Gusynin, V.A. Miransky, 
and Yu. A. Sitenko, Riv. del Nuovo Cim. {\bf N5} (1983) 1;
W. Bardeen, C.N. Leung, and S.T. Love, Nucl. Phys. 
{\bf B323} (1989) 493; V. A. Miransky, {\it Dynamical Symmetry 
Breaking in Quantum Field Theories} (World Scientific Co. Singarpore, 
1993);D.K. Hong and S.G. Rajeev, Phys. Rev. Lett. {\bf 64} (1990) 2475;
Phys. Lett. {\bf B240} (1990) 471.

\bibitem{georgi}A. Cohen and H. Georgi, Nucl. Phys. {\bf B 314} 
(1989) 7.



\end{thebibliography}
\end{document}